\documentclass[apjl]{emulateapj}
\usepackage{epsfig}
\begin{document}

\title{The effect of spiral structure on the measurements of the Oort 
constants}

\author{I.~Minchev and A.~C.~Quillen}
\affil{(Department of Physics and Astronomy,
University of Rochester), Rochester, NY 14627;
{\it iminchev@pas.rochester.edu, aquillen@pas.rochester.edu }
}

\begin{abstract}
We perform test-particle simulations in a 2D, differentially rotating 
stellar disk, subjected to a two-armed steady state spiral density 
wave perturbation in order to estimate the influence of spiral structure
on the local velocity field. By using Levenberg-Marquardt least-squares fit
we decompose the local velocity field (as a result of our simulations) into 
Fourier components to fourth order. 
Thus we obtain simulated measurements of the Oort 
constants, $A$, $B$, $C$, and $K$. We get relations between the Fourier 
coefficients and some galactic parameters, such as the phase angle of the 
Solar neighborhood and the spiral pattern speed.
We show that systematic errors due to the presence of spiral structure are 
likely to affect the measurements of the Oort constants.
Moderate strength spiral structure causes errors of order 5 km/s/kpc in $A$
and $B$.
We find variations of the Fourier coefficients with velocity dispersion,
pattern speed, and sample depth. For a sample at an average heliocentric 
distance of 0.8 kpc we can summarize our findings as follows:
(i) if our location in the Galaxy is near 
corotation then we expect a vanishing value for $C$ for all phase angles; 
(ii) for a hot disk, spiral structure induced errors for all Oort constants 
vanish at, and just inward of the corotation radius;
(iii) as one approaches the 4:1 Lindblad resonances $|C|$ increases 
and so does its variation with galactic azimuth;
(iv) for all simulations $|C|$, on average, is larger for lower stellar
velocity dispersions, contrary to recent measurements.
\end{abstract}

\section{Introduction}
Jan Oort proposed \citep{oort27a} and measured \citep{oort27b} the now 
called Oort constants $A$ and $B$, over seven decades 
ago by assuming that stars in the Galaxy moved on circular orbits. Later on 
his analysis was generalized to the non-axisymmetric case (e.g., 
\citealt{ogorodnikov32}) resulting in the definition of two more 
constants: $C$ and $K$. The importance of the Oort constants (OC) is in
their simple relation (at least in the case of vanishing random motions) 
to the galactic potential.

Many investigations have been conducted in the attempt to measure the OC. 
However, not all the results are in agreement (for a review see 
\citealt{kerr86}), which is mainly due to the absence of a complete proper 
motions catalog. Recent studies have failed to improve measurements of the
OC.

Another reason for the diversity among the values derived for the OC has 
been discussed by \citet{olling03}. The authors
present an effect which arises from the longitudinal variations of the 
mean stellar parallax caused by intrinsic density inhomogeneities. Together
with the reflex of the solar motion these variations create contributions
to the longitudinal proper motions which are indistinguishable from the OC
at $\le$ 20\% of their amplitude.  
\citet{olling03} (hereafter O\&D) measured the OC using proper 
motions from ACT/Tycho-2 catalog and corrected for the ``mode mixing" effect
described above, using the latitudinal proper motions.
They found that the OC vary 
approximately linearly with asymmetric drift (or age) from the young-stars 
values to those appropriate for old populations. This trend may result 
from perturbations induced by the Galactic bar or spiral structure.
In this paper we investigate the latter possibility. 

Deviations from the axisymmetric values of the OC
have not received a satisfactory interpretation although they are often
assumed to arise from a non-axisymmetric Galactic potential perturbation
such as spiral structure or/and the Galactic bar. 

Given their preeminent importance, we feel that a better understanding of
the effect of spiral perturbation on the measurements of the OC is
necessary. Spiral arms perturb the local velocity field corrupting local
measurements of the OC. Thus we look for the effect of spiral structure on
the OC $A,B,C,$ and $K$ and their deviation from the axisymmetric limit. 
We hope that this would give us some constraints on the spiral structure.
Our approach applies to the kinematics of the Local Standard of Rest (LSR),
while at the same time it constitutes a general analysis of a spiral density
wave perturbing a stellar disk.

Our simulation model is described in \S \ref{sec:model}. We present 
a derivation of the OC in \S \ref{sec:derive} and describe our numerical 
measurement technique in \S \ref{sec:measure}. Results are presented 
in \S \S \ref{sec:axi_hot}$-$\ref{sec:sample_depth}.
Discussion of observational  measurements of the OC is given in \S 
\ref{sec:observ} and summary follows in \S \ref{sec:sum}. 

\section{Numerical model}
\label{sec:model}

\subsection{Notation and units used}

For simplicity we work in units in which the galactocentric distance 
to the annulus in which stars are distributed is $r_0=1$, 
the circular velocity at $r_0$ is $v_0=1$, and the angular velocity
of stars is $\Omega_0 = v_0/r_0=1$. Because we assume a flat rotation curve
throughout this paper the initial circular velocity is $v_0=1$ everywhere.
One orbital period is $2\pi r_0 / \Omega_0 = 2\pi$. In our code time is in 
units of $1/\Omega$. The velocity vector of a star is $(u,v)$, where $u,v$
are the radial and tangential velocities in a reference frame rotating with
$v_0$. Consequently, the tangential velocity of a star in an inertial
reference frame is $v_{\phi}=v_0 + v$. We refer to the radial and tangential
velocity dispersions as $\sigma_u$ and $\sigma_v$, which are defined as the
standard deviations of $u$ and $v$, respectively.

The heliocentric distance, azimuth, radial and transverse velocities are 
denoted by $d, l, v_d$, and $v_l$, respectively. As it is commonly accepted,
we also refer to $l$ as the Galactic longitude. 
We write the longitudinal proper motion as $\mu_l$. 

The azimuthal wavenumber of the spiral wave, $m$, is an integer corresponding
to the number of arms. The gravitational potential perturbation amplitude
of the spiral density wave is denoted as $\epsilon_s$ and the pattern speed
is $\Omega_s$. The parameter $\alpha$ is related to the pitch angle of the 
spirals, $p$, with $\alpha = m \cot(p)$. $\alpha$ is negative for trailing
spirals with rotation counterclockwise. In this paper we only consider 
trailing spiral density waves.

Throughout this paper we refer to the dynamical Local Standard of Rest as 
LSR, which is a point at the position of the Sun moving with a circular 
velocity $v_0$. The Sun's velocity vector used in the derivation of the OC 
(\S \ref{sec:derive}) is defined as $(u,v_\phi)=(U_\odot, V_\odot)$. 

\subsection{Equations of motion}
\label{sec:ham}

We consider the $2D$ motion of a test particle in the mid-plane of a galaxy 
in an inertial reference frame. In plane polar coordinates ($r,\phi$) the 
Hamiltonian of a star can be written as 
\begin{equation}
\label{eq:ham}
H(r,\phi,p_r, p_{\phi},t)=H_0(r,p_r, p_{\phi}) + \Phi_1(r, \phi, t)
\end{equation}
The first term on the right hand side is the unperturbed axisymmetric 
Hamiltonian
\begin{equation}
H_0(r,p_r, p_{\phi}) =                            \ 
{p_r^2 \over 2} + {p_{\phi}^2 \over {2 r^2}} + \Phi_0(r)  
\end{equation}
where we assume the axisymmetric background potential due to the disk and 
halo has the form $\Phi_0(r)=v_0^2\log(r)$, corresponding to a flat rotation 
curve.

The imposed gravitational potential perturbation, $\Phi_1(r, \phi, t)$ is 
due to a spiral density wave and is discussed in the next section. 

The equations of motion derived from the Hamiltonian (eq. \ref{eq:ham}) are 
integrated particle by particle.

\subsection{Perturbation from spiral structure}

In accordance with the Lin-Shu hypothesis \citep{lin-shu}
we treat the spiral pattern as a small logarithmic perturbation to the 
axisymmetric model of the galaxy by viewing it as a quasi-steady density 
wave. We expand the spiral wave gravitational potential perturbation in 
Fourier components
\begin{equation}
\label{eq:phi}
\Phi_1(r, \phi, t)=\sum_m \epsilon_m
e^{i [\alpha \ln{r} - m(\phi-\Omega_s t - \phi_m )]}
\end{equation}
and its corresponding surface density
\begin{equation}
\label{eq:sigma}
\Sigma_1(r, \phi, t)=\sum_{m} \Sigma_m
e^{i [\alpha \ln{r} - m(\phi-\Omega_s t - \phi_m )]}
\end{equation}
We assume that the amplitudes, $\epsilon_m, \Sigma_m$, and the pitch angle
are nearly constant with radius. The strongest term for a two-armed spiral
is the $m=2$ term. Thus only the terms corresponding to $m=2$ are retained.
Upon taking the real part of eq. \ref{eq:phi} the 
perturbation due to a spiral density wave becomes
\begin{equation}
\Phi_1(r,\phi,t) = \epsilon_s \cos[\alpha \ln{r}-m(\phi-\Omega_s t-\phi_0)].
\end{equation}
The direction of rotation is with increasing $\phi$ and $\alpha< 0$ ensure 
that each pattern is trailing. 

The recent study of \citet{vallee05} provides a good summary of
the many studies which have used observations to map the Milky Way disk.
Cepheid, HI, CO and far-infrared observations suggest that the Milky Way
disk contains a four-armed tightly wound structure, whereas \citet{drimmel}
have shown that the near-infrared observations are consistent with a dominant
two-armed structure. A dominant two-armed and weaker four-armed structure, 
both moving at the same pattern speed, was previously proposed by 
\citet{amaral}. A combination of primary two- and weaker four-armed spiral
structures, moving at {\it different} pattern speeds was considered by 
\citet{mq06} as a possible heating mechanism. Here we consider a two-armed 
spiral density wave only and differ the four-armed case and the two+four 
arms combination to future investigations.

In general, an individual stellar orbit is affected by the mass distribution
of the entire galaxy. 
However, if tight winding of spiral arms is assumed only local gravitational 
forces need be considered. \citet{ma} find $kr$ for several spiral galaxies 
of various Hubble types to be $>$ 6, thus the tight-winding, or WKB 
approximation, is often appropriate. In the above expression $k$ is the 
wave-vector and $r$ is the radial distance from the Galactic Center (GC), 
related to the pitch angle $p$ through 
$\cot(p)=|kr/m|$. This gives $\alpha$ in terms of $k$ as $\alpha\approx|kr|$. 
In the WKB approximation the amplitude of the potential
perturbation Fourier component is related to the density perturbations in the 
following way
\begin{equation}
\epsilon_s \approx { - 2 \pi G  \Sigma_s r_0 \over |\alpha| v_0^2}
\end{equation}
(\citealt{B+T}). The above equation is in units of $v_0^2$. 
$\Sigma_s$ is the amplitude of the mass surface density of the two-armed 
spiral structure. 

\subsection{Resonances}
 
Of particular interest to us are the values of $\Omega_s$ which place the 
spiral waves near resonances. The Corotation Resonances (CR) occurs when the 
angular rotation rate of stars equals that of the spiral pattern.
Lindblad Resonances (LRs) occur when the 
frequency at which a star feels the force due to a spiral arm coincides with 
the star's epicyclic frequency, $\kappa$. As one moves inward or outward 
from the corotation circle the relative frequency at which a star encounters
a spiral arm increases. There are two values of $r$ for which this frequency
is the same as the epicyclic frequency and this is where the Outer Lindblad 
Resonance (OLR) and the Inner Lindblad Resonance (ILR) are located.

Quantitatively, LRs occur when $\Omega_s=\Omega_0 \pm \kappa/m$. 
The negative sign corresponds to the ILR and the positive - to the OLR. 
Specifically, assuming a flat rotation curve, for the 4:1 ILR 
$\Omega_s=0.65\Omega_0$, for the 4:1 OLR $\Omega_s=1.35\Omega_0$, for the 
2:1 ILR $\Omega_s=0.3\Omega_0$, and for the 2:1 OLR $\Omega_s=1.7\Omega_0$. 
Corotation of each spiral pattern occurs at $\Omega_s=\Omega_0$. 

\subsection{Simulation method}

To investigate the effect of a spiral density wave on the local velocity 
field we need to simulate a large number of stars.
In this paper we consider an initially cold or hot stellar disks as described
in \S \ref{sec:ic}.
We distribute particles in the annulus ($0.6r_0,1.4r_0$),
when simulating a cold disk, or ($0.4r_0,1.4r_0$) for the hot-disk case.
We reduce the computation time by recording the positions and
velocities of test particles 33 times per orbit for three orbits
after the initial seven rotation periods. Position and velocity vectors are
saved only if particles happened to be in the statistically important 
to us region ($0.8r_0,1.2r_0$). We require the final number of stars in the
case of the cold disk to be $5\times10^6$. For the hot-disk simulation
more particles are needed to beat the random motions: we require the output
to be $1.5\times10^7$. After utilizing the $m$-fold 
symmetry of our galaxy, we end up with a working sample of $~5\times10^5$
(cold disk) or $~1.5\times10^6$ (hot disk) stars in our fictitious Solar
Neighborhood (SN).

This way of creating additional particles is justified by the fact that 
a steady state (constant pitch angle and angular velocity) spiral 
density wave does not change the kinematics of stars once it has been grown.
As was demonstrated by \citet{mq06} test particles remain on the 
approximately closed orbits assumed during the growth of the spiral pattern 
unless the pattern speed places stars at an exact Lindblad resonance.

\subsection{Initial conditions}
\label{sec:ic}

The density distribution is exponential, $\Sigma(r)\sim e^{-r/r_\rho}$,
with a scale length $r_\rho = 0.38r_0$, consistent with the 2MASS photometry 
study by \cite{ojha01}. We would like to see how spiral structure changes 
the velocity field in the case of an initially cold or initially hot stellar
disk. To simulate the cold disk we send stars on circular orbits with the 
same initial azimuthal velocity $v_{\phi}=v_0+v=v_0$ consistent with a flat
rotation curve. For the hot disk we give an initial radial velocity 
dispersion in the form of a Gaussian distribution.
The Milky Way disk is known to have a radial velocity dispersion
which decreases roughly exponentially outwards: 
$\sigma^2_u\sim e^{-r/r_{\sigma^2}}$. In accordance with this we implement
an exponential decrease in the standard deviation of the radial velocity
dispersion of stars with radius, with a scale length $r_{\sigma^2}=0.45r_0$
\citep{lewis89}. We set $\sigma_u=40$ km/s at $r_0$.
% (see fig. \ref{fig:distr}). 
%The initial velocity ellipsoid ratio $\sigma^2_{\phi}/\sigma^2_u=0.5$. 
%However, when acted upon by the background potential this changes >>>>>?????
%In the latter expression $\sigma_v$ is
%the azimuthal velocity dispersion.  

The amplitude of the spiral density wave, $\epsilon_s$, is initially zero,
grows linearly with time at $0<t<t_1$ and is kept constant after $t=t_1=4$ 
rotation periods. Thus the spiral strength is a continuous function of $t$,
insuring a smooth transition from the axisymmetric to the perturbed state. 

Table \ref{table:par} summarizes the parameters for which simulations were 
run.

\subsection{Numerical accuracy}

Calculations were performed in double precision. We checked how energy  
was conserved in an unperturbed test run. The initial energy was compared to 
that calculated after 40 periods and the relative error 
was found to be $|\Delta E/E(0)|<8.48\times10^{-14}$. We also checked the 
conservation of the Jacobi integral $J$ in the presence of spiral structure.
The relative error found in this case was 
$|\Delta J/J(0)|<3.11\times10^{-12}$.

\section{Deriving the Oort constants}
\label{sec:derive}

\begin{figure*}
\epsscale{0.7}
\plotone{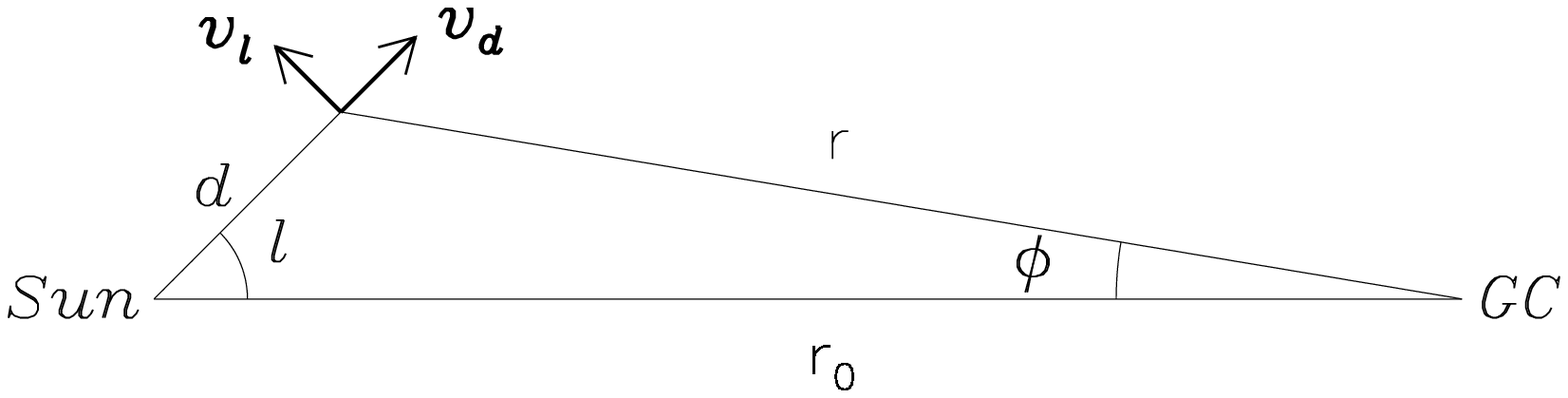}
\figcaption{
A diagram showing the geometry used in the derivation of the OC (see 
\S \ref{sec:derive}). The radial and transverse velocities vectors 
as seen from the Sun are indicated by $v_d$ and $v_l$, respectively. 
The galactocentric velocity vectors are omitted for clarity.
\label{fig:derive}
}
\end{figure*}

Consider a star moving in the gravitational potential described in \S 
\ref{sec:ham}. We assume that its velocity vector with respect to an 
inertial frame is $(u,v_{\phi})$ where $u$ is radial and $v_{\phi}=v_0+v$ 
is tangential. Let $u$ be positive toward the GC. We define $v_{\phi}$ to 
be positive in the same direction as the galactic rotation.
From the geometry of Fig. \ref{fig:derive} we can write the radial and 
transverse velocities of this star, $v_d$ and $v_l$ respectively, as seen by 
an observer sitting on the Sun, in terms of its 
galactocentric velocities $(u,v_{\phi})$ as
\begin{eqnarray}
\label{eq:vel}
v_l(l) = U_\odot \sin{l} - V_\odot \cos{l} + v_{\phi} \cos(l+\phi) \nonumber
- u \sin(l+\phi) \nonumber \\
v_d(l) = -U_\odot \cos{l} + V_\odot \sin{l} + v_{\phi} \sin(l+\phi)
+ u \cos(l+\phi)
\end{eqnarray}
where $l$ is the heliocentric azimuth, $r$ is the radius of the star from the
Galactic Center (GC), and $\phi$ is the angle between the star/GC and the
Sun/GC vectors. $(U_\odot,V_\odot)$ is the velocity vector of the Sun. 
Angle definitions lead to 
\begin{eqnarray}
\label{eq:angles}
\sin(l+\phi) &=& r \sin {\phi} \nonumber \\
d \cos{l} + r \cos{\phi} &=& r_0.
\end{eqnarray}
Using the laws of sines and cosines together with eq. (\ref{eq:angles}) we 
eliminate $\phi$ from eq. (\ref{eq:vel}) in favor of $d$, the heliocentric 
distance of the star, arriving at
\begin{eqnarray}
\label{eq:vel_2}
v_l(l) =   U_\odot \sin{l} - V_\odot \cos{l}         
         + {v_{\phi}\over r}(r_0\cos{l}-d) \nonumber
         - {u\over r} r_0\sin{l} \nonumber \\
v_d(l) = - U_\odot \sin{l} - V_\odot \cos{l} 
         + {u\over r}(r_0\cos{l}-d) \nonumber
         + {v_{\phi}\over r} r_0\sin{l}
\end{eqnarray}
To first order in $d$ or $\phi$,
\begin{eqnarray}
r \sim r_0 - d \cos l   \nonumber
\end{eqnarray}
and infinitesimals
\begin{eqnarray}
dr      &=&           -d \cos l \nonumber \\
d\phi &=& {d\over r_0} \sin l.  \nonumber
\end{eqnarray}
We now Taylor-expand $v_{\phi}$ and $u$ to first order about the Sun, 
$(r_0, \phi_0)$:
\begin{eqnarray}
v_{\phi}(r,\phi) &=& v_{\phi}(r_0, \phi_0)
    + {\partial v_{\phi} \over \partial r}(r_0, \phi_0) dr
    + {\partial v_{\phi} \over \partial \phi}(r_0, \phi_0) 
    d\phi \nonumber \\
u(r,\phi) &=& u(r_0, \phi_0)
    + {\partial u \over \partial r}(r_0, \phi_0) dr
    + {\partial u \over \partial \phi}(r_0, \phi_0) 
   d\phi. \nonumber
\end{eqnarray}
Assuming circular motion, the motion of the Sun implies that 
$v_{\phi}(r_0, \phi_0)  = V_\odot$ and $ u(r_0, \phi_0)  = U_\odot$. 
Using the above expressions for $v,r,dr,d\phi$ to first order in $d$ we can 
write the velocities $v_l,v_d$ as
\begin{eqnarray}
\label{eq:vel_oort}
v_l(l) &=& U_\odot\sin(l) - V_\odot\cos(l) + Ad\cos(2l) - Cd\sin(2l) + Bd\nonumber  \\
v_d(l) &=& -V_\odot\sin(l) - U_\odot\cos(l) + Ad\sin(2l) + Cd\cos(2l)  + Kd
\end{eqnarray}
where
\begin{eqnarray}
\label{eq:oc}
2A  &\equiv &   +{v_{\phi} \over r} - {\partial v_{\phi} \over \partial r} 
- {1\over r} {\partial u \over \partial \phi}  \nonumber          \\
2B  &\equiv &   -{v_{\phi} \over r} - {\partial v_{\phi} \over \partial r} 
+ {1\over r} {\partial u \over \partial \phi} \nonumber          \\
2C  &\equiv &   -{u \over r} + {\partial u \over \partial r}
- {1\over r} {\partial v_{\phi} \over \partial \phi} \nonumber       \\
2K  &\equiv &   +{u \over r} + {\partial u \over \partial r}
+ {1\over r} {\partial v_{\phi} \over \partial \phi}.
\end{eqnarray}
These equations are consistent with eq. (3) of O\&D in
slightly different notation.

We see from eqs. \ref{eq:vel_oort} that $A$ and $C$ can be determined from 
either the longitudinal proper motions $\mu_l\equiv v_l/d$ or radial 
velocities, $v_d$. When $A$ and $C$ are estimated from radial velocities 
their values are sensitive to the errors in the distances, whereas proper 
motions provide a distance independent way of measuring them. On the other 
hand, $B$ can only be measured from proper motions and $K$ only from radial
velocities. It is interesting to note that due to the uncertainty in the 
definition of an inertial coordinate system (\citealt{cuddeford94}, O\&D) 
$B$ is often estimated \footnote{The relation between the Oort ratio and 
the Oort constants is only justified in the case of well mixed, low 
velocity dispersion populations. Third order moments may contribute 
significantly if old stellar samples are used \citep{cuddeford94}.} 
from $A$ and the Oort ratio ${\sigma_v}^2/{\sigma_u}^2$
\begin{equation}
{{\sigma_v}^2\over{\sigma_u}^2} = {-B\over{A-B}}.
\end{equation}

If there is no streaming then
${\partial u \over \partial \phi}={\partial u \over \partial r}=
{\partial v_{\phi} \over \partial \phi}=0$, so that $C=K=0$, and $A$, $B$
reduce to the well known axisymmetric Oort constants
\begin{eqnarray}
\label{eq:oc_axi_hot}
2A_{\rm{axi}} &\equiv & +\left({v_{\phi}\over r} - 
{\partial v_{\phi} \over \partial r} \right)_{r_0},\nonumber\\
2B_{\rm{axi}} &\equiv & -\left({v_{\phi}\over r} + 
{\partial v_{\phi} \over \partial r} \right)_{r_0}.
\end{eqnarray}
If, in addition, the axisymmetric stellar disk is cold and the rotation
curve is flat we can define 
\begin{eqnarray}
\label{eq:oc_axi_cold}
2A_0 &\equiv & + {v_0 \over r_0}, \nonumber\\
2B_0 &\equiv & - {v_0 \over r_0}.
\end{eqnarray}

As seen in the derivation above, the OC are not really constant
unless they are measured in the SN. In fact, they may vary with
the position in the Galaxy $(r,\phi)$. The reason for variations of the
OC with Galactic radius, if we assumed the Galaxy were axisymmetric, is the
fact that the number density and velocity dispersions are functions of $r$. 
In the case of a non-axisymmetric Galaxy, due to spiral arms, bar,
etc., we would expect a variation with $\phi$ as well. 
Thus, the Oort constants have often been called the Oort {\it functions}.

\section{Numerical measurement of the Oort constants}
\label{sec:measure}

\begin{figure*}
\epsscale{0.7}
\plotone{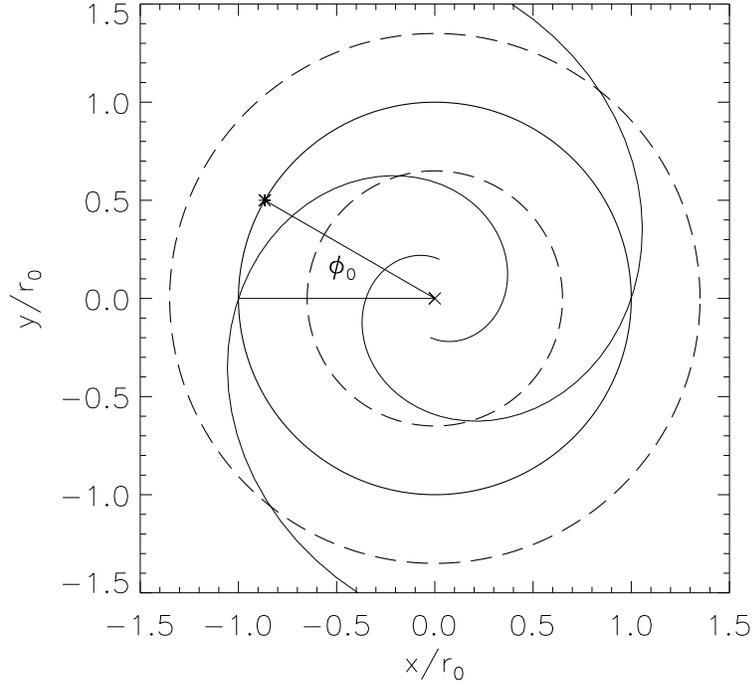}
\figcaption{
For a spiral density wave moving with $\Omega_s=\Omega_0$ the corotation
(solid) and 4:1 LRs (dashed) circles are shown. The phase angle of the SN
is given by $\phi_0$, the angle between the galactocentric lines passing 
through y=0 and the star symbol (representing the Sun). We treat $\phi_0$ 
as a free parameter in our simulations. In an inertial reference frame the 
rotation of both spirals and galaxy is in the clockwise direction. $\phi_0$
increases in the same direction. 
\label{fig:phi0}
}
\end{figure*}

First we move to a reference frame centered on $(r_0,\phi_0)$ where 
$\phi_0$ is the phase angle of the SN defined in fig. \ref{fig:phi0}. 
On the circle of radius $r_0$ we change the position of the SN between 
the convex and the concave spirals, thus treating $\phi_0$ as a free parameter.
We then select a sample of stars at a particular distance 
and annulus width from the point ($r_0,\phi_0$) which we identify with the 
LSR. The radial and transverse velocities of the SN stars, $v_d$ and $v_l$ 
respectively, as seen from the LSR are written in terms of the 
galactocentric velocities $(u,v_{\phi})$ (cf. eq. \ref{eq:vel_2}). Each 
annulus of stars is divided into 30 ``heliocentric" azimuthal bins with 
the Galactic longitude $l=0$ pointing toward the Galactic center. 
We obtain the functions $\bar{\mu}_l\equiv\overline{v_l(l)/d(l)}$ and 
$\bar{v}_d(l)/\bar{d}(l)$ where $\bar{\mu}_l$ is the longitudinal proper 
motion and the bars indicate average values. Hereafter we omit the bars.

\subsection{Fourier expansion of $\mu_l$ and $v_d/d$}
\label{sec:fourier}

We decompose $\mu_l\equiv v_l/d$ and $v_d/d$ into Fourier terms:

\begin{eqnarray}
\mu_l &=& c_{0l}+\sum_{i=1}^n (s_{il} \sin{il} + c_{il} \cos{il}). \nonumber \\
v_d\over d &=& c_{0d}+\sum_{i=1}^n (s_{id} \sin{id} + c_{id} \cos{id}).
\end{eqnarray}

By using Levenberg-Marquardt least-squares fit we obtain the expansion
coefficients up to $n=4$.
The OC $A$ is given by $c_{2l}$ and $s_{2d}$; $C$ is given by 
$-s_{2l}$ and $c_{2d}$; $B$ is obtained uniquely from $c_{l0}$; 
similarly, $K$ corresponds to $c_{d0}$.

\section{Hot axisymmetric disc}
\label{sec:axi_hot}

Before we look at the non-axisymmetric case we wish to examine the effect of 
an initially hot stellar disc on the measurement of the OC. As described 
above we use a background axisymmetric potential consistent with 
a flat rotation curve. If stars moved on circular orbits and we assumed a 
solar radius of $7.8$ kpc and a circular velocity of 220 km/s we would
measure $A=A_0\approx14.1$ km/s/kpc and $B=B_0=-A_0$. This would result in 
a value for the LSR rotation speed $\Omega_0=A-B=28.2$ km/s/kpc and slope 
of the rotation curve ${\partial v_\phi / \partial r}=-(A+B)=0$. 

However, in the case of an initial radial velocity dispersion $\sigma_u=40$
km/s, our simulations produced $A\approx12.1$ km/s/kpc and $B\approx-14$ 
km/s/kpc. In other words $B\approx B_0$, whereas the hot disk $A$ deviates 
from its axisymmetric value by $\sim2$ km/s/kpc. This is an interesting 
result for even if no perturbation is present we already have an error in 
measuring $A$. Consequently, we have $\Omega_0=A-B\approx26.1$ km/s/kpc and 
${\partial v_\phi /\partial r}=-(A+B)\approx1.9$ km/s/kpc. 
The inferred slope of the velocity rotation curve is no longer zero.

The reason for the observed decrease in $A$ is that our hot axisymmetric 
disk is described not by eqs. \ref{eq:oc_axi_cold} but by eqs. 
\ref{eq:oc_axi_hot}. Radial oscillations cause stars with home radii 
interior and exterior to $r_0$ to be observed at the LSR. Because the 
surface density and radial velocity dispersion are exponentially decreasing 
functions of radius, on average, these stars have a lower azimuthal velocity
than those with guiding radius of $r_0$. This causes a tail in the 
distribution of $v$ weighted toward negative values and is quantified by 
the asymmetric drift, defined as the difference between the LSR and the 
mean rotation velocity: $v_a\equiv v_0-\bar{v}_{\phi}$ \citep{B+T}. 
Obviously, the asymmetric drift is related to the radial velocity 
dispersion; it has been determined empirically that in the SN 
$v_a\simeq\sigma^2_u/k$ with $k=80\pm5$ km/s \citep{db98} ($k$ here should 
not be confused with the spiral wave number). In our notation 
$v_a=-\bar{v}$ so we can rewrite eqs. \ref{eq:oc_axi_hot} as
\begin{eqnarray}
\label{eq:oc_va}
2A_{\rm{axi}} &=& 2A_0-{v_a \over r_0} +
       \left({\partial v_a \over \partial r}\right)_{r_0} \nonumber \\
2B_{\rm{axi}} &=& 2B_0+{v_a \over r_0} +
       \left({\partial v_a \over \partial r}\right)_{r_0}.
\end{eqnarray}
We numerically estimated the values of the asymmetric drift and its slope 
and found $v_a(r_0)=16.9$ km/s and ${\partial v_a(r) /\partial r}=-1.9$ 
km/s/kpc, respectively. These values, together with the numerically derived 
$A_{\rm{axi}}\approx12.1$ and $B_{\rm{axi}}\approx-14$ specified above were 
found to satisfy eqs.
\ref{eq:oc_va}. The last two terms on the right side of the second of these
equations happen to almost completely cancel out, resulting in 
$B_{\rm{axi}}\approx B_0$.

The deviations from the cold axisymmetric OC (small for $B$ and larger for 
$A$) estimated from our hot axisymmetric disk are in very good agreement 
with the analytical results of \cite{kuijken91} and O\&D (their eq. 12).
The asymmetric drift inferred from $v_a\simeq\sigma^2_u/k$, given a radial
velocity dispersion $\sigma_u=40$ km/s, is 20 km/s $\pm1.3$. Even though
this is marginally consistent with our numerically calculated value 
$v_a=16.9$ km/s, here consistency is not to be expected since the constant 
$k$ was determined empirically for the SN and thus may reflect 
non-axisymmetric effects or incomplete mixing.

Eqs. \ref{eq:oc} were derived only to first order in the Taylor expansion of 
the velocities. Higher order terms might become important for deep samples.
In our simulations we calculate the OC from a sample of test particles 
centered at $d=800$ pc from the LSR and an annulus width of 200 pc. Since
our expressions for the OC are only valid in the limit of $d\rightarrow0$,
we need to evaluate the OC in several distance bins and extrapolate to 
$d=0$. We performed such an extrapolation for the hot axisymmetric disk
discussed above and found that in the range $0<d<1.4$ kpc the variation
in $A$ and $B$ was less than $\sim0.5$ km/s/kpc. 
However, when spiral structure is included, the variations with sample depth
could be significant, especially for the cold disk case. \S 
\ref{sec:sample_depth} is devoted to this problem.

The errors in $A$ and $B$ induced by the axisymmetric drift in the hot 
disk need to be kept in mind when considering the hot population
simulations with included spiral structure perturbation, 
presented in the next section.

\section{The effect of spiral structure}
\label{sec:results_sp}

We now describe how the presence of spiral perturbation causes the OC
to deviate from their axisymmetric values.
In all of our simulation runs we used a two-armed spiral density wave. 
To look for variation of the OC with the phase of the SN, $\phi_0$,
(see \S \ref{sec:measure} for definition of $\phi_0$) we split the arc
between two spiral arms into 50 SN disks. We estimate the OC
in each of these disks as described in \S \ref{sec:fourier}. 

It is interesting to see how pattern speeds placing the background 
stars near resonances affect the local velocity field. Possible locations
of the LSR have been proposed to be near the 4:1 ILR \citep{qm05} or near
the CR \citep{lepine03,mish02,mish99}. We examine two particular cases:
\newline
(i) an initially cold stellar disk and 
\newline
(ii) an initially hot disk ($\sigma_u=40$ km/s).
\newline
To find what errors result in measuring the OC when spiral structure
perturbs the disk, in the figures below we plot the difference between
$A$, $B$ estimated from a simulation run with an imposed spiral
perturbation and $A$, $B$ resulting from an axisymmetric disk.
Of course, for a cold disk (circular orbits) $A=A_0$ and $B=B_0$ but, as we
found in the previous section, an axisymmetric hot disk yields a substantial
decrease in $A$ and a small increase in $B$. 
Therefore, we calculate the spiral structure induced errors in $A$ and $B$
differently, depending on whether the disk is initially cold or hot:  
For case (i) we plot $\Delta A\equiv A-A_0$ and for case (ii) 
$\Delta A\equiv A-A_{\rm{axi}}$. Similar expressions hold for $\Delta B$. 

First we present our results for the variation of the OC with phase angle and
pattern speed in a more general way. In \S \S \ref{sec:4:1ilr} and 
\ref{sec:CR} we concentrate on the 4:1 ILR and CR, respectively.
Table \ref{table:par} summarizes the parameters used for each run.  

\subsection{Cold stellar disk}
\label{sec:cold}

\begin{figure*}
\epsscale{1.0}
\plotone{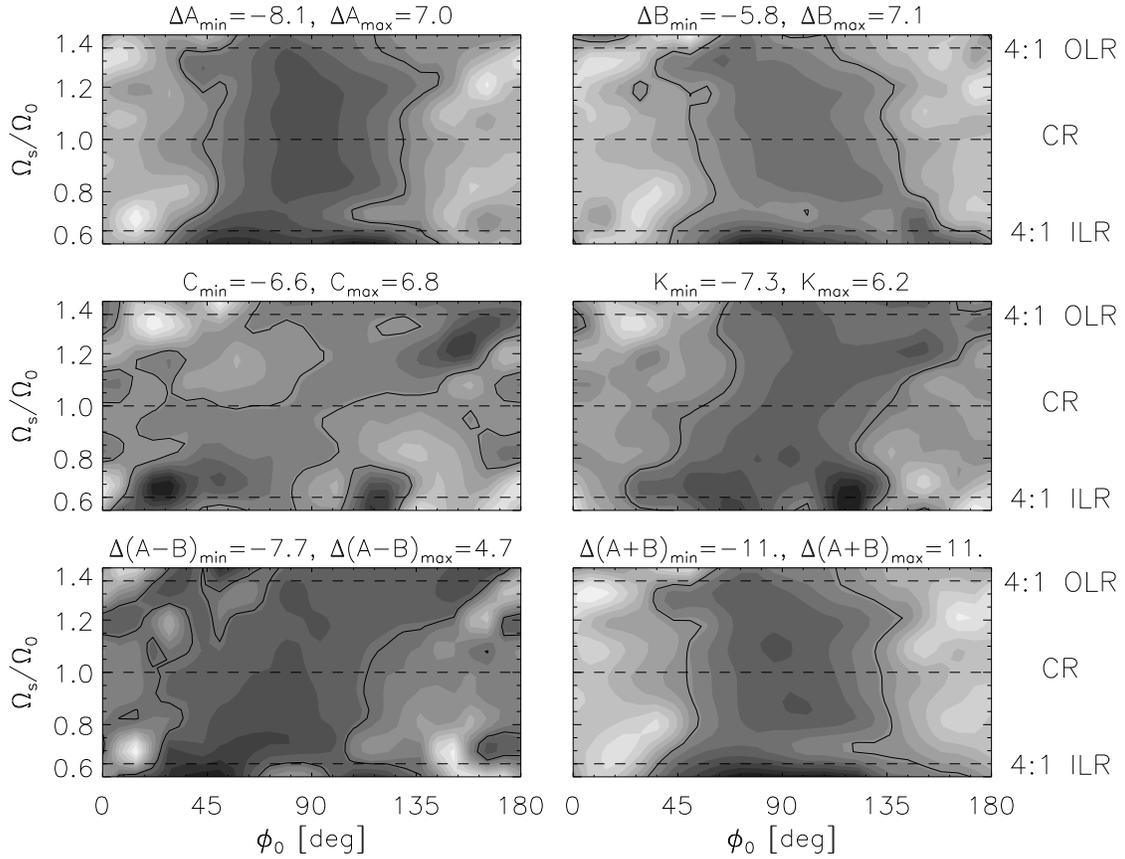}
\figcaption{
Contours show the spiral structure induced errors in the values of the OC 
and combinations thereof, as a function of the Sun's phase angle, $\phi_0$, 
and the pattern speed, $\Omega_s/\Omega_0$. A two-armed spiral perturbation 
with strength $\epsilon_s=1.0$ is imposed on an initially cold stellar disk.
The phase angle varies between the convex spiral arm at 
$\phi_0=0^\circ$, and the concave one at $\phi_0=180^\circ$. Minimum and 
maximum values are given above each panel. Darker regions correspond to 
larger values. The zero contour level in each panel is indicated by a solid
line. Dashed lines show the occurrence of resonances. It is interesting to 
note that $C$ is nearly zero at the corotation resonance for all angles. 
We note that the y-axis can also be interpreted as a variation of the 
galactic radius $r/r_0$ (see \S \ref{sec:cold}). Spiral structure parameters
used in our simulations are found in table \ref{table:par}.
\label{fig:cont_u00}
}
\end{figure*}

We first investigate the effect of spiral structure perturbing a cold
stellar disk. In fig. \ref{fig:cont_u00} we plot contours of $\Delta A$, 
$\Delta B$, $C$, $K$, $\Delta(A-B)$, and $\Delta(A+B)$, against the LSR
phase angle, $\phi_0$, and the pattern speed, $\Omega_s$. We define 
$\Delta(A\pm B)\equiv \Delta A\pm\Delta B$. Each horizontal line corresponds
to a simulation with a different spiral angular velocity, $\Omega_s$. 
Here $\Delta A$ and $\Delta B$ are the deviations from the axisymmetric 
values $A_0$ and $B_0$. Darker regions correspond to larger values.
Minimum and maximum values are indicated above each panel. The zero contour 
level in each panel is indicated by a solid line. We change the LSR angle
between the convex arm at $\phi_0=0^\circ$ and the concave one at 
$\phi_0=180^\circ$ (see fig. \ref{fig:phi0}). Dashed lines show the locations 
of CR and 4:1 LRs. 

There is another interpretation of the y-axis of each panel in fig. 
\ref{fig:cont_u00}. Since logarithmic spirals are self-similar and the 
velocity rotation curve is flat, changing the pattern speed at the same 
radius is equivalent to changing the radius and keeping the spiral angular 
velocity the same. Thus, we note that fig. \ref{fig:cont_u00} could be 
perceived as the variation of the OC with respect to radius, $r/r_0$. For
example, assuming the LSR were located at the CR (or $\Omega_s=\Omega_0$)
the vertical axis varies between $r=0.6 r_0$ and $r=1.4 r_0$. Note, however, 
that this interpretation of the y-axis of fig. \ref{fig:cont_u00} does not 
reflect the slope change in the exponential surface density distribution 
which would result with the change of radius.

$\Delta A$ and $\Delta B$ are maximized midway between the spiral arms 
(top left and right panels in fig. \ref{fig:cont_u00}). There is some
structure in $A$ and $B$ as one approaches the 4:1 ILR and contours are 
skewed toward the convex (small $\phi_0$) or concave (large $\phi_0$) arm, 
respectively.

The Oort's $C$ is directly related to the radial streaming caused by
the presence of the spiral arms. Its variation with $\phi_0$ and $\Omega_s$ 
in the cold disk case is shown in the left column, second row panel in fig. 
\ref{fig:cont_u00}. Structure appears to be symmetric across the diagonals 
with the most notable exception of the clump at 
$(\phi_0,\Omega_s)\approx(120^\circ, 0.7\Omega_0)$. 
An interesting result is that $C\approx0$ at the CR for all angles. The
variation of $C$ with $\phi_0$ becomes more prominent as
one moves from the CR to each of the 4:1 LRs. That is where minimum and 
maximum values are reached: $C_{\rm{min}}=-6.3$ km/s/kpc at the point 
$(\phi_0,\Omega_s)\approx(20^\circ, 1.35\Omega_0)$ and $C_{\rm{max}}=6.5$ 
km/s/kpc at $(\phi_0,\Omega_s)\approx(25^\circ,0.7\Omega_0)$. 
We also note that all structure in this panel seems shifted toward small 
$\phi_0$, or the convex arm, by about $5^\circ$.

If the LSR were located near the CR then $C$ would not provide any 
information about the spiral structure, regardless of the phase angle.
On the contrary, $A$ and $B$ show marked variation with galactocentric 
azimuth along the CR. However, if in addition we happened to be at 
$\phi_0\approx45^\circ$ or $\phi_0\approx135^\circ$, then we would measure 
values for $A$, $B$, and $C$ (and even $K$) close to those expected for an 
axisymmetric galaxy. 

We discuss how an initially hot stellar disk is affected by spiral structure
in the next section.

\subsection{Hot stellar disk}
\label{sec:hot_sp}

In \S \ref{sec:axi_hot} we showed that a hot axisymmetric disk already 
causes miss-measurements of $A$ and $B$. We expect this result to be 
reflected in the case of a hot stellar disk with an imposed spiral 
perturbation. To separate only the effect of the perturber on $A$ and $B$ 
we need to correct for the asymmetric drift induced error. In the following 
figures we plot $\Delta A\equiv A-A_{\rm{axi}}$ and 
$\Delta A\equiv B-B_{\rm{axi}}$ as discussed at the beginning of \S 
\ref{sec:results_sp}.

\begin{figure*}
\epsscale{1.0}
\plotone{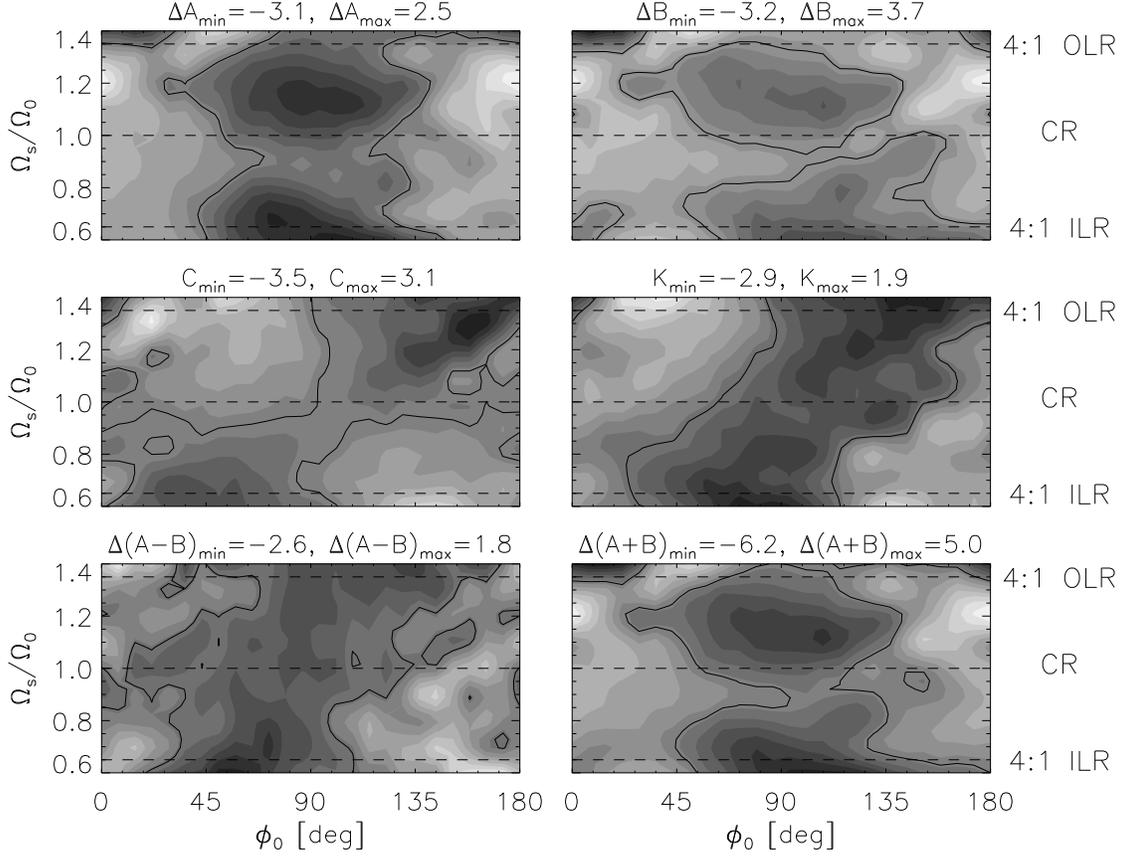}
\figcaption{
Same as Fig. \ref{fig:cont_u00} but for an initially hot stellar disk. The
initial radial velocity dispersion is $\sigma_u=40$ km/s. The asymmetric
drift induced errors in $A$ and $B$ are subtracted so that the plots reflect
the effect of the spiral structure only (see beginning of \S 
\ref{sec:results_sp}).
The hot population causes the central regions of the cold $\Delta A$ and
$\Delta B$ to split into two islands with a line of symmetry along 
$\Omega_s/\Omega_0\approx0.9$. In addition to $C\approx0$ along the corotation
resonance found for the cold disk, here $\Delta A$, $\Delta B$, and $K$ are also 
nearly zero for all $\phi_0$. 
It is interesting that for the hot disk the symmetry line of all contour
plots has shifted from the CR to $\Omega_s/\Omega_0\approx0.9$.
If the Sun is located at, or just inward of, the corotation radius then 
measurements of the Oort constants, using a high velocity dispersion population, 
would provide no information about the spiral structure.
Deviations from the axisymmetric values of all constants is decreased as a 
direct consequence of the initially hot stellar population. 
\label{fig:cont_u40}
}
\end{figure*}

Fig. \ref{fig:cont_u40} shows the effect of a two-armed spiral density wave
on an initially hot axisymmetric background disk. The simulation setup is 
identical to fig. \ref{fig:cont_u00} with the exception of an initial radial
velocity dispersion $\sigma_u=40$ km/s. We found in \S \ref{sec:axi_hot} 
that this initial setup caused an asymmetric drift of $v_a=16.9$ km/s when 
non-axisymmetric perturbation was absent. 

For the cold disk, maximum $\Delta A$ and $\Delta B$ values were attained 
in the central regions of their contour plots (top panels of fig. 
\ref{fig:cont_u00}). With the increase of velocity dispersion both of these
have split into two islands, with lines of symmetry at 
$\Omega_s/\Omega_0\approx0.9$ (fig. \ref{fig:cont_u40}). Along this line, 
and even along the CR (see fig. \ref{fig:o10}), $\Delta A$ and especially 
$\Delta B$ are very close to zero. As a direct consequence of the effect 
of the initially hot stellar population, the spiral structure induced 
errors in $A$ and $B$ have decreased by factors of $\sim3$ and $\sim2$, 
respectively.

Since the contour plots of $\Delta A$ and $\Delta B$ in fig. 
\ref{fig:cont_u40} exhibit quite similar morphology they cancel out in 
$\Delta(A-B)$ and add up in $\Delta(A+B)$ (bottom two panels in fig. 
\ref{fig:cont_u40}). As a result, the $\Delta(A-B)$ contour plot is very 
similar to its cold disk counterpart, whereas the contours of $\Delta(A-B)$
resemble those of $\Delta A$ and $\Delta B$.    

The value of $C$ along the CR line is still close to zero and again, the 
most variation with galactic azimuth takes place near the 4:1 LRs. 
However, similarly to $\Delta A$ and $\Delta B$, minimum and maximum values 
have decreased in magnitude, by a factor of $\sim2$ for $C$. 
This was expected in view of the fact that with the increase of the stars' 
random motions, the response to spiral structure must decrease.
The prominent peak seen in the cold $C$ contour plot at 
$(\phi_0,\Omega_s)\approx(120^\circ,0.7\Omega_0)$ (fig. \ref{fig:cont_u00}) 
has completely disappeared in the hot disk plot (fig. \ref{fig:cont_u40}).   

In addition to $C\approx0$ along the CR line found for the cold disk, here 
$\Delta A$, $\Delta B$, and $K$ are also nearly zero for all $\phi_0$.
It is interesting that for the hot disk the symmetry line of all contour
plots has shifted from the CR to $\Omega_s/\Omega_0\approx0.9$.
If the Sun is located at, or just inward of, the corotation radius
then measurements of the OC, using a high velocity dispersion population,
would provide no information about the spiral structure.

In their observational measurements O\&D find that $C$ varies approximately 
linearly with color, $\bv$, with $C\approx0$ km/s for early type and 
$C\approx-10$ km/s for the later stars (red giants). They also found 
similar variation with asymmetric drift $v_a$, where $C\approx-10$ km/s 
for $v_a=18$ km/s. Early type stars are younger and have small velocity 
dispersion. On the other hand, high $\bv$ suggests older stellar population 
and thus higher $\sigma_u$, thus larger $v_a$. This contradicts our 
intuitive expectation, as well as our numerical results, that the radial 
streaming induced by spiral structure becomes less important as velocity 
dispersion increases. 

We next look at the variation of the OC with Sun's phase angle
at a fixed spiral pattern speed. 

\subsection{Near the 4:1 ILR with spiral pattern}
\label{sec:4:1ilr}

\begin{figure*}
\epsscale{1.0}
\plotone{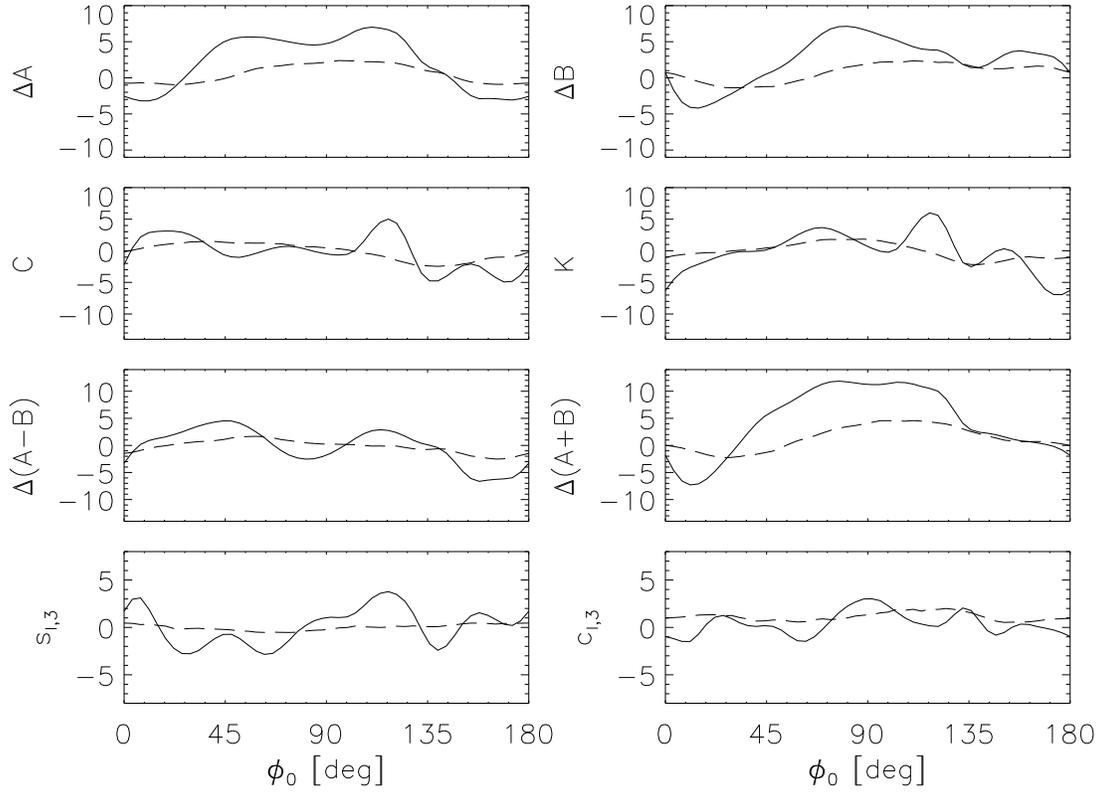}
\figcaption{
Variation of the OC and combinations as a function of the LSR phase angle, 
$\phi_0$, at the same spiral pattern speed, $\Omega_s=0.6\Omega_0$
(just inside the 4:1 ILR).
Solid and dashed lines represent slices from fig. \ref{fig:cont_u00} 
(cold disk) and fig. \ref{fig:cont_u40} ($\sigma_u=40$ km/s), respectively.  
The units of the y-axis are km/s/kpc. Deviations from axisymmetry decrease 
with increasing velocity dispersion in all panels since
random motions take precedence over the spiral structure perturbation.
\label{fig:o06}
}
\end{figure*}

Variation of the OC and their combinations as a function of LSR phase angle, 
$\phi_0$, at the same spiral pattern speed $\Omega_s=0.6\Omega_0$, 
or just inside the 4:1 ILR, is shown in fig. \ref{fig:o06}.
Solid and dashed lines represent slices from fig. \ref{fig:cont_u00} 
(cold disk) and fig. \ref{fig:cont_u40} ($\sigma_u=40$ km/s), respectively. 
The units of the y-axis are km/s/kpc. Deviations from axisymmetry (zero
y values) decrease with increasing velocity dispersion in all panels since
random motions take precedence over the spiral structure perturbation:
both less structure and decrease of amplitude is observed for the hot disk
(dashed lines).

Strong, two-armed spirals have been shown to create square-shaped orbits 
near their 4:1 ILR \citep{cont86}. The orientation of the orbits interior 
to the 4:1 ILR is such that they support the spiral structure. On the other 
hand, at guiding radii exterior to the 4:1 ILR orbits of stars fail to 
support the spiral structure. We expect this behavior to be reflected in our
results for the OC.

\begin{figure*}
\epsscale{1.0}
\plotone{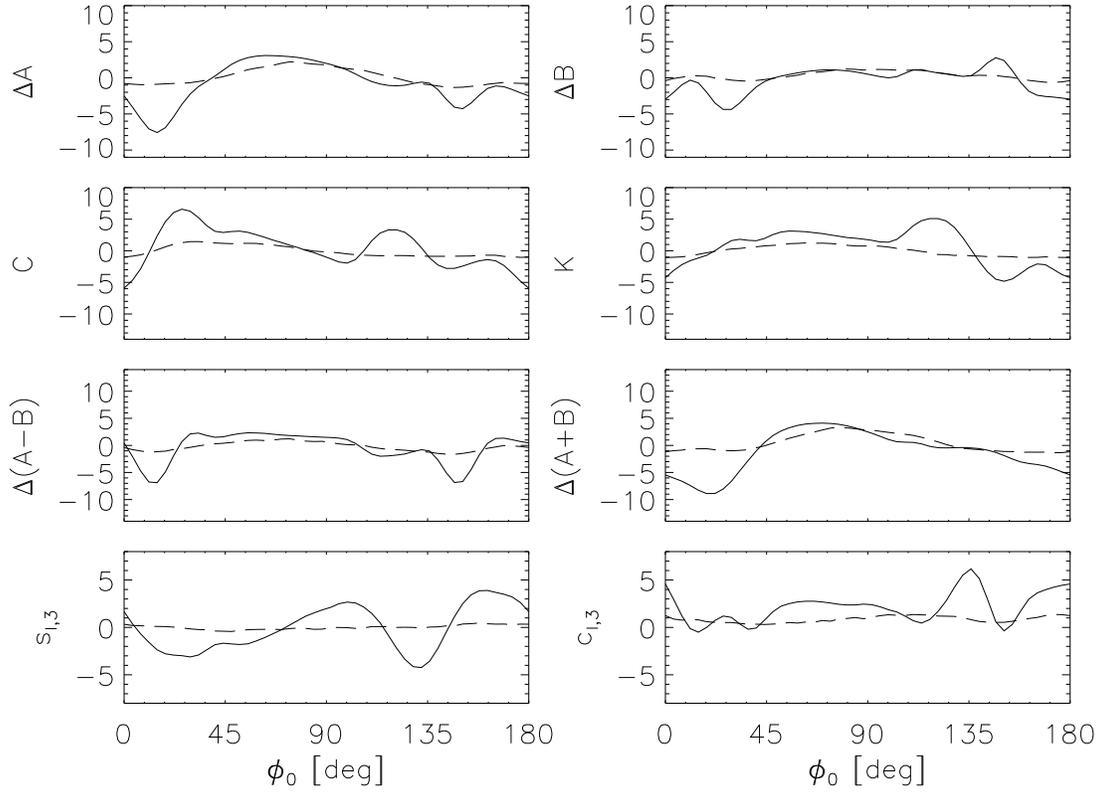}
\figcaption{
Same as fig. \ref{fig:o06} except that $\Omega_s = 0.7\Omega_0$ (just outside 
the 4:1 ILR). 
\label{fig:o07}
}
\end{figure*}

On the outside of the 4:1 ILR, $\Omega_s=0.7\Omega_0$, we observe a
significant change in the variation of the OC for the cold disk 
(solid lines in fig. \ref{fig:o07}). However, the OC resulting from the hot 
disk (dashed lines in fig. \ref{fig:o07}) are almost identical to those in
fig. \ref{fig:o06}. Again, the variations present in the cold disk values are
reduced by the random motions of the stellar population.  

The bottom two panels in figs. \ref{fig:o06} and \ref{fig:o07} show the 
third-order coefficients in the Fourier expansion of $\mu_l$, 
$s_{l,3}$ and $c_{l,3}$ (see \S \ref{sec:fourier}).
These are quite large for the cold disk and almost zero for the hot one.
As with $C$, non-zero $s_{l,3}$ and $c_{l,3}$ are indicators of 
non-axisymmetry.

\subsection{At the CR with spiral pattern}
\label{sec:CR}

\begin{figure*}
\epsscale{1.0}
\plotone{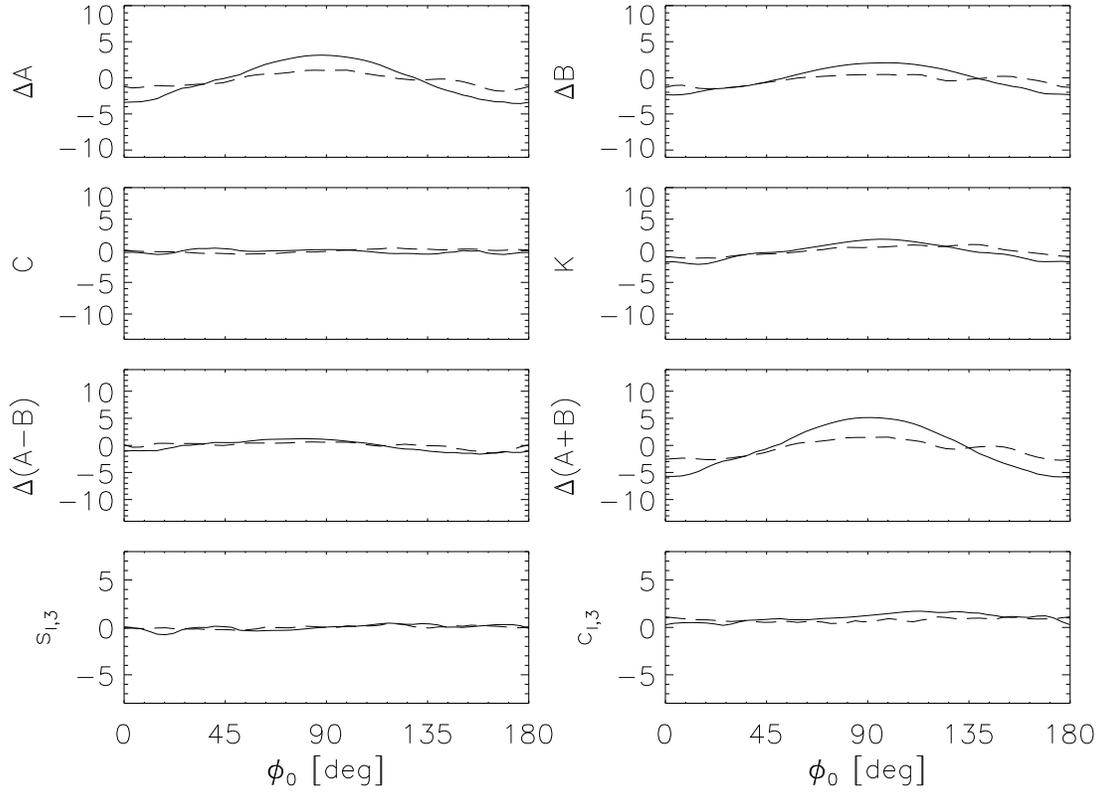}
\figcaption{
Same as figs. \ref{fig:o06} and \ref{fig:o07}
but with $\Omega_s = \Omega_0$ (CR). 
We observe almost the same variation in $\Delta A$ and $\Delta B$. 
This gives rise to $\Delta A-B\approx0$ but large slope of the galactic
rotation curve, ${\partial v_\phi / \partial r}=-(A+B)$.
It is very interesting that $C\approx0$ for all phase angles for both
cold and hot disks. If LSR is located 
at or near the corotation resonance \citep{lepine03,mish02,mish99} then $C$
will provide no information about the spiral structure. 
\label{fig:o10}
}
\end{figure*}

Unlike the prominent variations of the OC with phase angle near the 
4:1 LRs, at the CR we 
observe only mild effects (fig. \ref{fig:o10}). The hot-disk OC values 
still deviate from axisymmetry less than their cold-disk counterparts, 
however the difference is not as pronounced as it is near the 4:1 LRs 
(cf. figs. \ref{fig:o06} and \ref{fig:o07}). 

$\Delta A$ and $\Delta B$ exhibit similar variation with $\phi_0$.
As a result, $\Delta(A-B)$ is very small (bottom left in fig. \ref{fig:o10}).
On the other hand, the inferred slope of the galactic velocity rotation 
curve, $\partial v_{\phi}/\partial r=-(A+B)$, deviates from zero except 
for particular phase angles. It is very interesting that
$C\approx0$ for all phase angles for both cold and hot disks, as mentioned 
in the discussion of figs. \ref{fig:cont_u00} and \ref{fig:cont_u40}.  
If the Sun is located near the CR, this behavior of $C$ would provide no
information on the spiral structure. The third order Fourier expansion 
coefficients $s_{l,3}$ and $c_{l,3}$ are very close to zero for both cold 
and hot disks.
\newline

We ran a simulation with a four-armed spiral structure with a pattern speed
placing background stars at the CR. Our results were similar to the 
two-armed case. 
\newline

In all of our simulations involving a spiral density wave perturbation the 
absolute values of $C$ for the initially cold disk were larger than those 
for the hot disk for most values of the phase angle, $\phi_0$, and pattern 
speed $0.6<\Omega_s/\Omega_0<1.4$. 
This shows that it is not possible to reproduce the observational measurements
of O\&D with the spiral structure alone. 

\section{Variation with sample depth} 
\label{sec:sample_depth}

\begin{figure*}
\epsscale{0.85}
\plotone{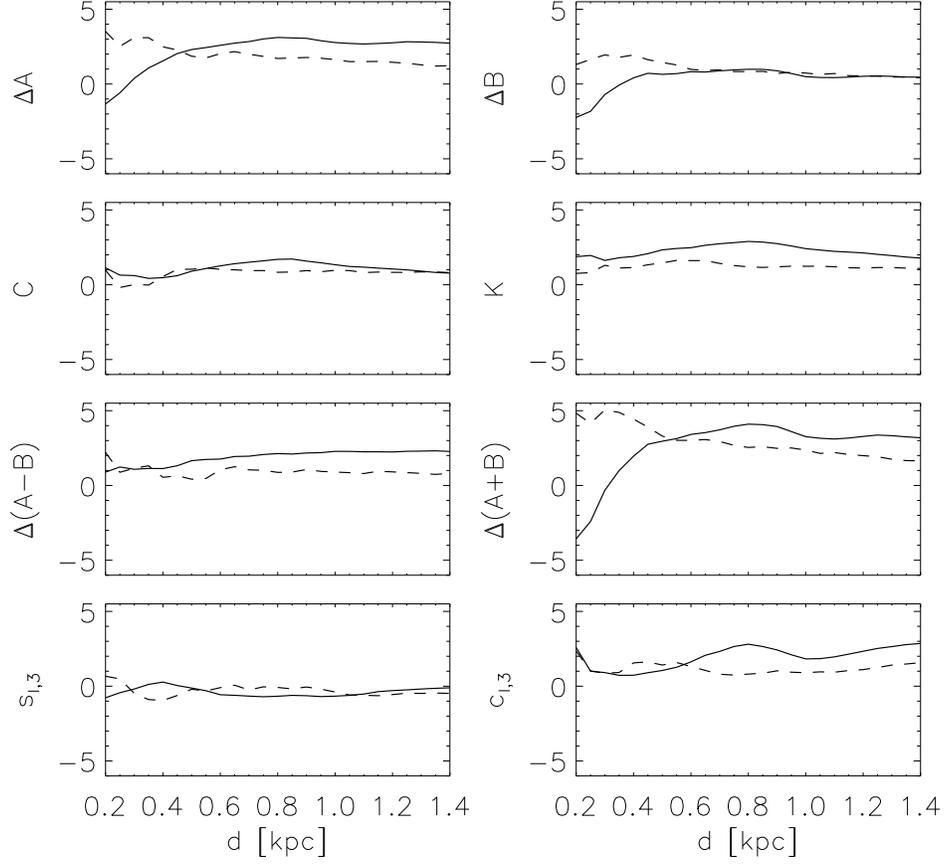}
\figcaption{
Variation of the OC with sample depth for $\Omega_s = 0.7\Omega_0$ (just 
outside the 4:1 ILR) and a phase angle $\phi_0=68^\circ$ (that is, the Sun is
$68^\circ$ ahead of the convex arm). Solid and dashed lines show the results
from simulations with an initially cold and hot disks, respectively. The 
x-axis shows the change of the average heliocentric distance of the sample of
stars used to estimate the OC.
The y-axis units are km/s/kpc. Note that in all previous figures the Fourier 
coefficients presented were estimated at an average distance of $d=0.8$ kpc. 
At large $d$ the inferred slope of the rotation curve, 
${\partial v_\phi / \partial r}=-(A+B)$, is negative whereas for small $d$ 
it is positive. This is in very good agreement with the shape of the actual 
wiggle in the average $v_\phi(r)$ induced by the spiral structure at this 
location, shown in the first panel of fig. \ref{fig:v_vs_r}.
\label{fig:o07_19}
}
\end{figure*}

\begin{figure*}
\epsscale{0.85}
\plotone{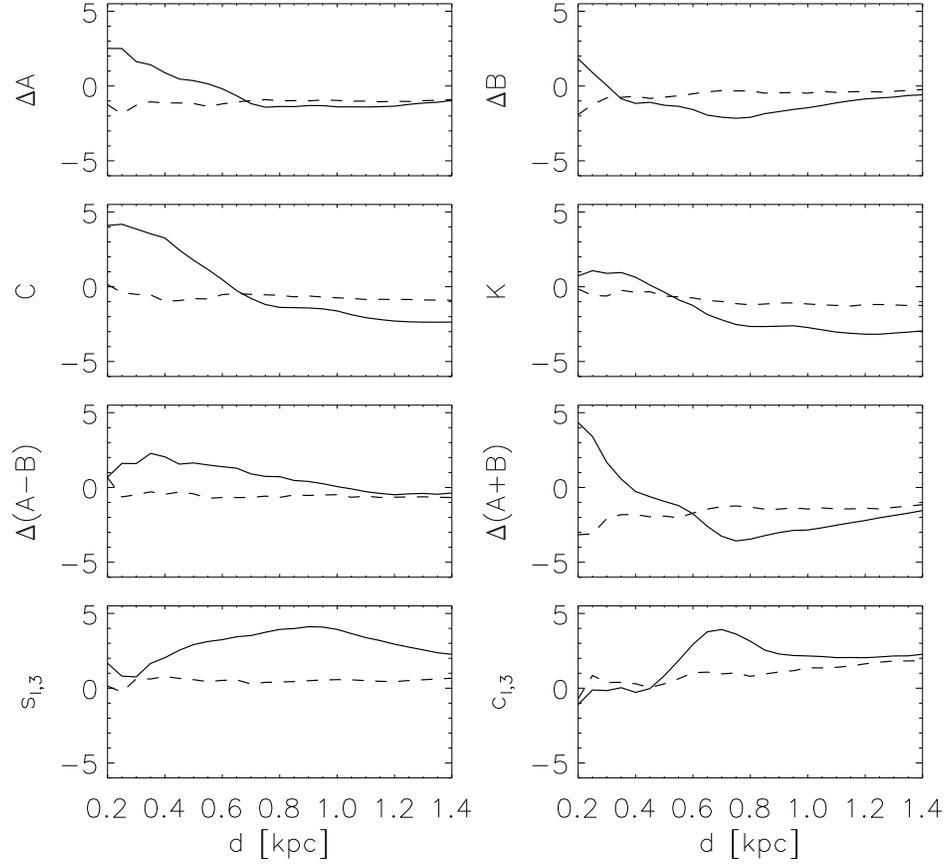}
\figcaption{
Same as fig. \ref{fig:o07_19} but for a phase angle $\phi_0=162^\circ$.
Here the inferred rotation curve has the opposite behavior to fig. 
\ref{fig:o07_19}: it is rising for large $d$ and declining if closer samples 
are considered. Again, this is in very good agreement with the spiral structure
induced wiggles in the initially flat $v_\phi(r)$, shown in the 
second panel of fig. \ref{fig:v_vs_r}.
\label{fig:o07_45}
}
\end{figure*}

For all of the simulations with included spiral structure discussed above 
we estimated the OC from a sample at an average heliocentric distance 
$d=0.8$ kpc and annulus width of 0.2 kpc. 
We found in \S \ref{sec:axi_hot} that for a hot axisymmetric disk changing
$d$ between 0.2 and 1.4 kpc caused almost no variations of $A$ and $B$.
However, when the spiral density wave perturbation is included this need 
not be the case because there are strong spatial variations in the mean 
velocity field. Following the same approach we estimated the OC in
different heliocentric distance bins in the range $0.2<d<1.4$ kpc. 
The results for $\Omega_s=0.7\Omega_0$, or just outside
the 4:1 ILR, and phase angles $\phi_0=68^\circ$ and $\phi_0=162^\circ$ are 
presented in figs. \ref{fig:o07_19} and \ref{fig:o07_45}, respectively.
Solid and dashed lines show the results from simulations with initially 
cold and hot disks, respectively. The y-axis units are km/s/kpc. 
In fig. \ref{fig:o07_19} we observe a large variation with sample depth for
the cold $\Delta A$ and $\Delta B$ for $d<\sim500$ pc. Due to the similar 
variation in $\Delta A$ and $\Delta B$ mainly the combination $\Delta (A+B)$
is affected, resulting in a different value for the inferred slope of the 
galactic rotation curve, depending on the heliocentric distance of the 
sample considered. The cold disk yields much larger variations of $A$ and 
$B$ (about 30\% for $A$) with sample depth compared to the hot values 
(about 10\%).

We also looked at the variations of the OC with average distance from the 
Sun at a phase angle $\phi_0=162\circ$, which places the SN near to the 
concave (leading) arm. 
In fig. \ref{fig:o07_45}, $\Delta A$ and $\Delta B$ are found to increase 
with decreasing $d$, which is the opposite to the trend observed in fig. 
\ref{fig:o07_19}. The result again is a large variation for $\Delta (A+B)$ 
and smaller for $\Delta (A-B)$. However, at this phase angle all other OC
(resulting from the cold disk), including the third order coefficients, 
are greatly affected by the change in $d$.

\begin{figure*}
\epsscale{0.85}
\plotone{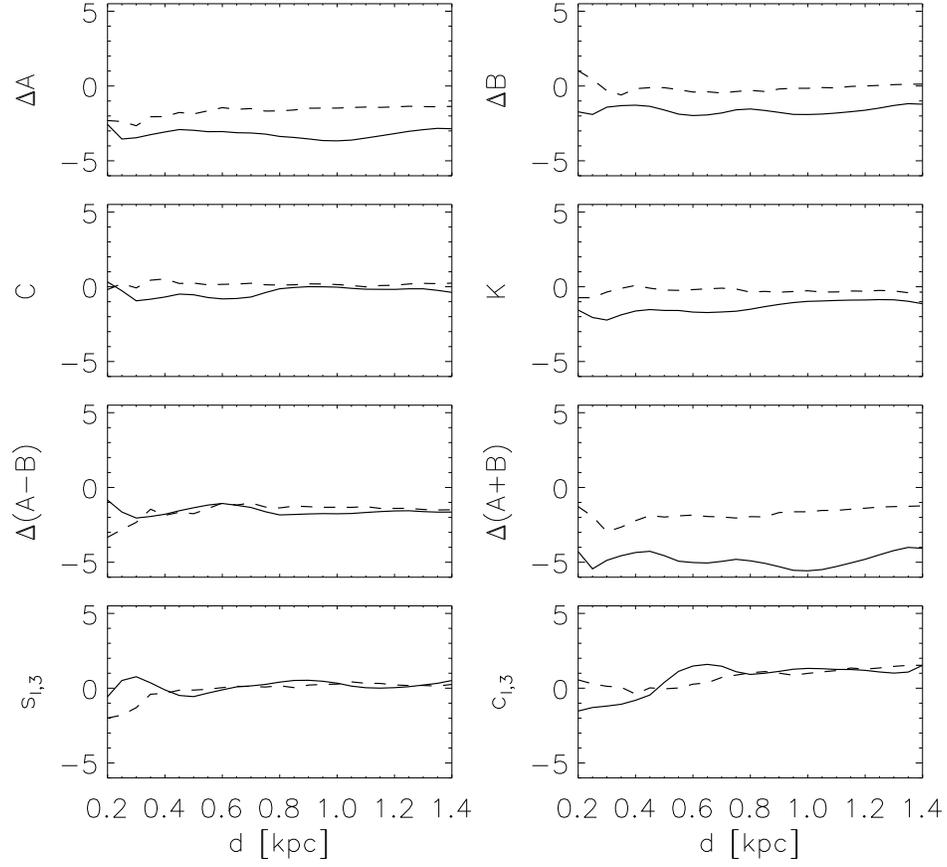}
\figcaption{
Same as figs. \ref{fig:o07_19} and \ref{fig:o07_45} but for a pattern speed
$\Omega_s = \Omega_0$ (corotation resonance). The phase angle is 
$\phi_0=162^\circ$, same as in fig. \ref{fig:o07_45}. At this pattern speed 
the variation of the usual OC with sample depth for both cold and hot disks 
is only of the order of $\sim1$ km/s/kpc, however the coefficient of the
$\cos{3l}$ Fourier term varies by more than $\sim2$ km/s/kpc.
\label{fig:o10_45}
}
\end{figure*}

We also plotted the variation of the OC with $d$ for a simulation with 
$\Omega_s=\Omega_0$ and $\phi_0=162^\circ$ (fig. \ref{fig:o10_45}). 
At this pattern speed the usual OC vary with sample depth, for 
both cold and hot disks, by $\sim1$ km/s/kpc, however the coefficient 
of the $\cos{3l}$ Fourier term changes by more than $\sim2$ km/s/kpc.

The following argument was found to shed some light on the question: 
Why do the cold OC vary so much with the depth of the sample considered 
whereas the hot ones show only mild variations? A problem with the 
determination of the OC addressed by O\&D is the effect of a non-smooth 
mean velocity field. Because of local anomalies in the Galactic potential 
caused by spiral arms, for example, the streaming field would exhibit small 
scale 
oscillations on top of the underlying smooth field. This would result in an
oscillating mean velocity field, giving rise to significant higher order terms 
in the Taylor expansion of the velocities. O\&D have simulated this effect 
in their fig. 2 for an axisymmetric streaming velocity. They find that for 
wiggles of wavelength 2 kpc and amplitude of only 2\% on an otherewise 
smooth rotation velocity, depending on the depth of the stellar sample 
considered, the measured $A$ and $B$ could deviate by as much as 30\%. 
A true representation of the smooth background potential would be achieved 
only at distances larger than the wavelength of the small scale oscillation.

\begin{figure*}
\epsscale{0.5}
\plotone{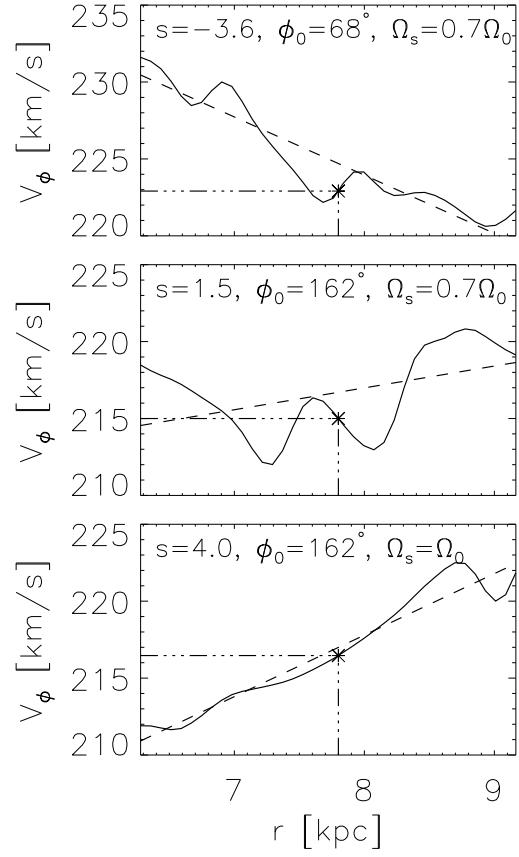}
\figcaption{
Wiggles in the average tangential velocity $v_\phi(r)$, induced by a two-armed 
spiral structure propagating in a cold stellar disk.
The phase angle and pattern speed are indicated in each panel. A linear fit is 
performed (dashed lines) with the slope values given by ``$s$". The Sun's 
position is at $r_0=7.8$ kpc, indicated by the star symbol. The top, middle, and
bottom panels correspond to the simulations giving rise to figs. \ref{fig:o07_19},
\ref{fig:o07_45}, and \ref{fig:o10_45}, respectively.
\label{fig:v_vs_r}
}
\end{figure*}

To test whether it is a non-smooth streaming field that gives rise to the
vehement variation of the cold OC with sample depth, we plotted the 
tangential velocity as a function of galactic radius in fig. 
\ref{fig:v_vs_r}. The x-axis extends from $(r_0-1.4)$ to $(r_0+1.4)$ kpc 
which covers the maximum sample depth considered in figs. 
\ref{fig:o07_19}$-$\ref{fig:o10_45}. The top, middle, and bottom panels 
represent the spiral structure induced wiggles in the mean tangential 
velocity field, $v_\phi(r)$, corresponding to the cold disk values of the 
OC given in figs. \ref{fig:o07_19}, \ref{fig:o07_45}, and \ref{fig:o10_45}, 
respectively. We fitted a line to each of the plots in fig. \ref{fig:v_vs_r} 
(dashed lines) with the slopes indicated by ``$s$" in the figure. The 
position of the Sun is given by the star symbol at $r_0=7.8$ kpc. 

As discussed by O\&D, estimating the OC when a non-smooth streaming field 
is present would give different results for different sample depths. Indeed,
considering fig. \ref{fig:o07_19}, at $d\approx0.2$ kpc the inferred slope
of the rotation curve is ${\partial v_\phi /\partial r}=-(A+B)\approx3.5$, 
it decreases quickly to $\approx-3.5$ at $d\approx0.6$ kpc and remains 
approximately constant for greater distances. Now looking at the top panel 
of fig. \ref{fig:v_vs_r} we see that the same behavior is apparent here. 
The slope is positive if a sample of stars close to the Sun is considered. 
As $2d$ becomes larger than the wavelength of the velocity curve oscillation 
(about 1 kpc), the slope of the smoother background velocity field is 
recovered: $s=-3.6$ km/s/kpc. For the same plot we fitted a line to 
$v_\phi(r)$ of stars with $|r-r_0|<0.8$ kpc and found a value of $s=-4$, 
consistent with the maximum achieved in fig. \ref{fig:o07_19} at 
$d\approx0.8$ kpc (panel showing $\Delta(A+B)$). 

The middle panel of fig. \ref{fig:v_vs_r} corresponds to the cold-disk OC 
values of fig. \ref{fig:o07_45}. Here the phase angle is $\phi_0=162^\circ$
which causes the spiral structure induced wiggle in the velocity curve to 
have the opposite slope (in the range $|r-r_0|<1.4$) to the top panel of 
fig. \ref{fig:v_vs_r}. Again, we find a very good agreement with the inferred
value for the slope given by ${\partial v_\phi /\partial r}=-(A+B)$ in fig. 
\ref{fig:o07_45} and the line fit to $v_\phi$ shown in fig. 
\ref{fig:v_vs_r}. At small distances we find $s<0$; $v_\phi(r)$ is 
approximately flat for $0.4<d<0.6$ kpc; finally, at an average $d=1.4$ kpc 
from the Sun both methods estimate $s\approx1.5$ km/s/kpc. 

Lastly, the bottom panel of fig. \ref{fig:v_vs_r} shows $v_\phi(r)$ for the
cold disk at the corotation radius and a phase angle $\phi_0=162^\circ$.
Here $v_\phi(r)$ appears much smoother compared to the simulation with 
$\Omega_s=0.7\Omega_0$. As in the two cases above, the variation in 
$v_\phi(r)$ is accurately reflected by $\Delta(A+B)$ in the corresponding 
fig. \ref{fig:o10_45}. Note that as in the middle panel (same phase angle), 
the induced slope in the region $|r-r_0|<1.4$ is positive. 

It is important to realize that the slope of the ``large" scale velocity 
field we measure for distances greater than $\sim1$ kpc is induced by the 
asymmetric drift at that particular phase angle and is just a larger scale 
wiggle on the otherwise smooth, mean $v_\phi(r)$. By asymmetric drift 
here we refer to the local, spiral structure induced radial streaming, 
causing stars exterior or interior to $r_0$ to be seen at $r_0$. Unlike the
asymmetric drift of a hot axisymmetric disk (see \S \ref{sec:axi_hot}), when
a non-axisymmetric perturbation is present $v_a$ and its gradient depend 
not only on the galactic radius but also on the azimuth. 
%For our spiral wave we can roughly expect that 
%$v_a<0$ and $\partial v/\partial r<0$ for a phase angle placing the SN near the
%convex arm and the oposite behaviour when near the concave one. 
Since we start with initially cold orbits the strong forcing of
the spirals near a resonance (4:1 ILR or CR in this case) could cause both 
positive and negative values for the derivative of the asymmetric drift, 
${\partial v_a /\partial r}$, depending on the phase $\phi_0$. Consequently, the 
mean $v_\phi(r)$ could be declining or rising as shown in fig. \ref{fig:v_vs_r}. 
In fact, even $v_a<0$ values are possible in a cold galaxy as is the case 
of the top panel of fig. \ref{fig:v_vs_r} (since $v_\phi(r_0)>v_0=220$ km/s). 
This was generally found to occur
at locations closer to the convex arm for both $\Omega_s=0.7\Omega_0$ and 
$\Omega_s=\Omega_0$. In reality a negative asymmetric drift has never been
observed. A possible reason for this could be our 
phase with respect to the spiral arms. Another explanation could be the 
fact that stars are actually born with some velocity dispersion. 
We simulated a disk with an 
initial $\sigma_u=10$ km/s and found that $v_a<0$ was still possible.  
The spiral structure induced velocity curve wiggles at 
$\phi_0=162^\circ$ shown in the middle ($\Omega_s=0.7\Omega_0$) and bottom 
($\Omega_s=\Omega_0$) panels of fig. \ref{fig:v_vs_r} correspond to positive
asymmetric drift and a phase angle lagging the concave spiral arm by 
$18^\circ$, both consistent with what has been proposed to be the case for 
the SN.

Examining figs. \ref{fig:o07_19}$-$\ref{fig:o10_45} we find the largest 
variation of $C$ with sample depth for $\phi_0=162^\circ$ and 
$\Omega_s=0.7\Omega_0$ (fig. \ref{fig:o07_45}). We expect this 
behavior to arise from spiral structure induced oscillations in the first
two terms on the right side of the third of eqs. \ref{eq:oc} (assuming the
third one can be neglected), similarly to the variations in $A$ and $B$ 
being the result of an oscillating local average velocity field, 
$v_\phi(r)$. The interplay of these terms at different 
average distances from the Sun was found to account for the variation of $C$ 
with sample depth at this location.

Due to the strong variations of the cold-disk OC with sample depth, $d$, we 
cannot extrapolate to $d=0$ as we did for the hot axisymmetric disk. Instead,
observational measurements of the OC should be compared to non-axisymmetric
models at the same distance from the Sun and for the same velocity dispersions.

We conclude that low velocity dispersion stars (bluer samples) 
would be good tracers for the local variation of the mean velocity field. 
For the hot disk the dependence of the OC on sample depth is usually not 
significant. The results of this section show that we cannot neglect the
variations of the OC with sample depth for cold populations. It is remarkable, 
however, that even when higher order terms are present, the cold-disk
$A$ and $B$ provide a good estimate for the slope of the mean velocity field.

\section{The Oort constants from observation}
\label{sec:observ}

It is interesting to compare our results to the observationally deduced
values for the OC.
From the then available proper motions and radial velocities,
\citet{oort27b} measured $A\approx19$ and $B\approx-24$ km/s/kpc.
Since then many measurements have been done, mainly of $A$ and $B$. 
The review of \citet{kerr86} yields the average values $A=14.5\pm1.3$ and 
$B=-12.0\pm2.8$ km/s/kpc. \citet{kuijken91} reviewed the observational
constraints on $C$ and $K$ and concluded that $C=0.6\pm1.1$ and 
$K=-0.35\pm0.5$ km/s/kpc. Thus these are zero within the errors as in the
case of an axisymmetric Galaxy.

The release of the Hipparcos catalog (ESA 1997) made it possible to 
re-analyze the local velocity field.  
In their work with the Hipparcos Cepheids, \citet{feast97} found 
$A=14.8\pm0.8$ and $B=-12.4\pm0.6$ km/s/kpc. These values are very
close to the \citet{kerr86} averages stated above but have smaller errors. 
\citet{mignard00} derived $A=11.0\pm1.0$ and $B=-13.2\pm0.5$ km/s/kpc 
for early type dwarfs, and $A=14.5\pm1.0$ and $B=-11.5\pm1.0$ km/s/kpc
for the distant giants. This increase in $|A|$ and $|B|$ with average age is 
consistent with our results for the case of a spiral perturbation.
At particular phase angles (mostly near the spiral arms), figs. 
\ref{fig:o06}$-$\ref{fig:o10} show larger values for hotter populations. 

\citet{torra00} analyzed a large sample of nearby O and B stars from the
Hipparcos catalog which they complemented with distance estimates from
Str{\"o}mgren photometry \citep{hauck98} and radial velocities. They 
analyzed the local velocity field in the context of the Gould Belt. The 
latter object is a ring-like structure inclined at $16-22^\circ$ with respect
to the Galactic plane, extending to about 600 pc from the Sun. 
Their inferred values for stars in the range $0.6<d\leq2.0$ kpc are 
$A=13.0\pm0.7$, $B=-12.1\pm0.7$, $C=0.5\pm0.8$, and $K=-2.9\pm0.6$ km/s/kpc. 
Using O and B Hipparcos stars, \citet{lindblad97} found $C=0.8\pm1.1$ and 
$K=-1.1\pm0.8$ km/s/kpc for $r>0.6$ kpc (or outside the Gould belt), in 
agreements with \citet{torra00}. \citet{CTG94}, using B6-A0 stars in the 
range $d<1.5$ kpc, found $K\approx-1$ km/s/kpc in agreement with 
\citet{torra00} and \citet{lindblad97}. However, their inferred value for 
$C$ is clearly negative: $C=-8.8\pm1.1$ km/s/kpc, and is consistent
with the ``phase mixing" corrected $C$ derived by O\&D. 

We could not account for the last result (large negative $C$ from blue stellar 
samples) by perturbing a stellar disk with 
spiral structure. The rest of the above observational measurements of the OC
are easily reproduced with the influence of a spiral perturbation for
a variety of pattern speeds and phase angles (cf. figs. 
\ref{fig:o06}$-$\ref{fig:o10}).

However, all of the above results could suffer from the ``mode mixing"
problem, pointed out by O\&D. They argue that non-axisymmetry is the 
dominant origin of the observed differences of the OC between stellar 
samples. If the reason was non-equilibrium effects then one would expect 
a somewhat erratic behavior instead of the clean trends seen in O\&D's 
fig. 9, left-hand panel. The authors concluded that sample depth is also 
unlikely to account for these changes. We agree with this conclusion for 
the case of the high velocity dispersion population. However, as we showed
in the previous section, spiral structure propagating in a cold disk 
could cause marked variations in all OC as a
function of sample depth, $d$.

It should be kept in mind that the various references presented here could 
be subject to various selection effects. For example, the results presented
by \cite{kerr86} are mostly based on Northern hemisphere proper motion and
radial velocity surveys. This causes a bias in the OC towards smaller $|A|$
and $|B|$ \citep{olling98}. In addition, before the release of the Hipparcos
catalog Oort's $B$ was obtained in a non-inertial reference frame and 
should be viewed with caution. 

%Distant stellar populations, e.g. Cepheids, O/B stars \citep{feast97}, 
%result in higher values for the tangential component $V_\odot$ compared to
%values derived from nearby Hipparcos stars \citep{db98}. The diference is 
%$\sim5$ km/s. In our simulations we assume the Sun moves with the velocity
%of the LSR, $(u, v_\phi)=(0,v_0)$ (thus we use SN and LSR interchangingly) 
%and measure the OC with respect to this point. In reallity, since the 
%Galactic potential is not exactly known, averaging over the tangential 
%velocities in the SN may not be the same as the tangential velocity of the 
%LSR. Spiral structure could cause small-scale ocsillations in the velocity
%of the colder populations. 
%Solar motion is found to vary with the group of stars under consideration.

Lastly, the solar motion is another source of error in estimating the OC.
The motion of the Sun with respect to the mean velocity vector of a
sample of stars leads to $\cos l$ and $\sin l$ terms for the mean radial
velocity and proper motions as a function of Galactic longitude.
These $n=1$ terms (see \S \ref{sec:fourier}) have been fit to our 
numerically generated stellar sample, 
so the mean velocity field for each
simulated measurement has been removed before the simulated OC
measurement. One problem confronting observational surveys is that
the mean velocity vector with respect to the Sun may in fact depend on
the stellar sample. By removing the $n=1$ terms in our simulated
measurements we have bypassed this problem, however our simulated
sample is not affected by distance biases. As pointed out by O\&D an
additional problem is that variations in the mean distance of the
sample with Galactic longitude can lead to aliasing or ''mode mixing" 
as well as difficulty in
measuring the mean velocity vector of the sample.

\section{Summary and Discussion}
\label{sec:sum}

The importance of the Oort constants is in their simple relation to the 
galactic potential, in the case of vanishing random motions. The circular 
orbits have velocity $r\partial\Phi/\partial r=v^2_\phi=(A_0-B_0)^2$.
If we could find the causes for the deviations, $\Delta A$ and $\Delta B$, 
from the ``true" values, $A$ and $B$, then 
we would provide a direct constraint on the Galactic potential.

In this paper we have investigated the effect of spiral structure on the 
measurements of the OC $A$, $B$, $C$, and $K$. We have performed 
test-particle simulations with an initially cold or hot stellar disk and 
an imposed two-armed spiral density wave perturbation. 
Variations of simulated measurements of the OC with pattern 
speed, galactic azimuth, and sample depth have been explored. 

We find that systematic errors due to the presence of spiral structure 
{\it are} likely to affect the measurements of the Oort constants.
Moderate strength spiral structure causes errors of order 5 km/s in $A$ and 
$B$. Axisymmetric, high velocity dispersion disks also yield errors in the 
simulated measurements of Oort's $A$ (as a result of the asymmetric drift),
but they are smaller, of order 2 km/s (see \S \ref{sec:axi_hot}).

If the Sun is located near the corotation resonance then, 
regardless of the phase angle, we would measure $C\approx0$ and thus get no 
information about the spiral structure. If, in addition, we constrain 
$\phi_0\approx45^\circ$, or $135^\circ$, all OC would have values 
within the current measurement error. As the Lindblad resonances are neared,
$C$ varies strongly with phase angle and has an increased value. 

In \S \ref{sec:sample_depth} we investigated the effect of sample depth on
the measurements of the OC. We found that for an initially cold disk,
spiral structure can give rise to marked variations of all OC with the 
average heliocentric distance of the sample. 
For $A$ and $B$ this results from an oscillating average $v_\phi(r)$,
whereas $C$ is sensitive to variations in the radial 
velocity and its gradient.
We find that for phase angles placing the Sun near the convex arm negative
values for the axisymmetric drift are possible for low velocity dispersion
stars. Unlike the $\sigma_u$ induced asymmetric drift in a hot axisymmetric
disk, in a cold disk it is the spiral structure that causes it. 
It is possible the reason for never observing $v_a<0$ could be our proximity
to the concave spiral arm.

Our result for the variation of $C$ with velocity dispersion is 
distinctly different than the observational measurement of $C$ done by O\&D. 
Whereas in our simulations the absolute value of $C$ decreases 
with increasing velocity dispersion, the opposite trend is observed in 
fig. 6 by O\&D. In the same paper the authors suggest that it is possible
that the Galactic bar is responsible for this behavior. 
Since orbits change orientation at the 2:1 OLR of the bar, at this location
the bar is expected to most strongly distort the local kinematics.
As discussed by O\&D, stars from inside the OLR should produce $C<0$.
If the Sun's location is just outside the OLR then we would sample more
stars giving rise to negative $C$ with increasing sample depth.
To test this possibility, O\&D analyzed only stars brighter than $V_T=10.5$
which yielded a decrease in the ``phase mixing" corrected value of $C$ from
$\sim-10$ to $\sim-7.5$ km/s. Another way to try to explain it is to 
consider the combined effect of spiral structure and bar perturbations.
Other possibilities are the effects from minor mergers, or those due to a 
triaxial halo \citep{kuijken94}.
Future work should aim at understanding this unusual behavior for $C$.

\acknowledgements
We would like to thank Jason Nordhaus for helpful comments. 
Support for this work was in part
provided by National Science Foundation grant ASST-0406823,
and the National Aeronautics and Space Administration
under Grant No.~NNG04GM12G issued through the Origins of Solar Systems 
Program.

{}

\clearpage

\clearpage

%------------------------------------------------------------------------------
\begin{deluxetable}{lccccc}
\tablewidth{0pt}
\tablecaption{Parameters describing simulations\label{table:par}}
\tablehead{
\colhead{Figures}                       &
\colhead{$\phi_0$}                      &
\colhead{$\Omega_s/\Omega_0$}           &
\colhead{$\sigma_u$ [km/s]}             &
\colhead{$\bar{d}$ [kpc]}               &
}
\startdata
\ref{fig:cont_u00}
   & $0-180^\circ$ &  $0.6-1.4$   &   0     & 0.8\\
\ref{fig:cont_u40}
   & $0-180^\circ$ &  $0.6-1.4$   &   40    & 0.8\\
\ref{fig:o06}(solid/dashed)
   & $0-180^\circ$ &  0.6         &   0/40  & 0.8\\
\ref{fig:o07}(solid/dashed)
   & $0-180^\circ$ &  0.7         &   0/40  & 0.8\\
\ref{fig:o10}(solid/dashed)
   & $0-180^\circ$ &  1.0         &   0/40  & 0.8\\
\ref{fig:o07_19}(solid/dashed)
   & $68^\circ$    &  0.7         &   0/40  & $0.2-1.4$\\
\ref{fig:o07_45}(solid/dashed)
   & $162^\circ$   &  0.7         &   0/40  & $0.2-1.4$\\
\ref{fig:o10_45}(solid/dashed)
   & $162^\circ$   &  1.0         &   0/40  & $0.2-1.4$\\
\ref{fig:v_vs_r}(top/mid/bottom)
   & $68^\circ$/$162^\circ$/$162^\circ$ &  0.7/0.7/1.0  &   0     & 
\enddata
\tablecomments{
Spiral pattern parameters corresponding to the simulations shown in the 
Figures. For all simulations there are three parameters that we do not change:
(i) the spiral perturbation strength is $\epsilon_s=-0.01$, given in units of
$v_0^2$, the velocity of a star in a circular orbit at $r_0$;
(ii) the parameter $\alpha=-6$, sets the pitch angle of the spiral arms, 
$p$, as $m \cot(p) = \alpha$; and 
(iii) $m=2$ since we only consider two-armed spirals.
$\phi_0$ is the angle between the galactocentric lines passing through the
solar neighborhood and the intersection of the convex spiral arm and the 
circle $r_0$; $\phi_0=0$ on the line crossing the 
convex arm. The pattern speed, $\Omega_s$ is in units of $\Omega_0 = v_0/r_0$. 
The radial velocity dispersions is given by $\sigma_u$.
Finally, the average distance of the ``heliocentric" bins is indicated by 
$\bar{d}$. 
}
\end{deluxetable}

%------------------------------------------------------------------------------

\end{document}